\documentstyle{llncs}

\newcommand\nc{\newcommand} \nc\rnc{\renewcommand}

\nc\nn{\newenvironment} \nc\nt{\newtheorem}

\newstytheorem{Claim}{\bf}{\it}[theorem]{Claim}

\rnc\a{\alpha} \rnc\b{\beta} \rnc\d{\delta} \nc\e{\varepsilon}
\nc\g{\gamma} \nc\r{\rho} \rnc\t{\tau}  \nc\U{\Upsilon}
\nc\cA{\cal A} \nc\bx{\bar{x}} \nc\by{\bar{y}}

\nc \Cclose{C_{\mbox{\footnotesize close}}}
\nc \copen {c_{\mbox{\footnotesize open}}}
\nc \Dclose{D_{\mbox{\footnotesize close}}}
\nc \dclose{d_{\mbox{\footnotesize close}}}
\nc \dopen {d_{\mbox{\footnotesize open}}}
\nc \dmax  {d_{\mbox{\footnotesize max}}}
\nc \dmin  {d_{\mbox{\footnotesize min}}}

\nc \false {\mbox {\em false}}
\nc \true  {\mbox {\em true}}
\nc \undef {\mbox {\em undef}}

\nc\CD {\mbox{Clear\-Deadline}}
\nc\CT {\mbox{CT}}
\nc\Dir{\mbox{Dir}}
\nc\DL {\mbox{Deadline}}
\nc\GS {\mbox{Gate\-Status}}
\nc\SC {\mbox{Signal\-Close}}
\nc\SD {\mbox{Set\-Deadline}}
\nc\SO {\mbox{Signal\-Open}}
\nc\TS {\mbox{Track\-Status}}
\nc\WT {\mbox{Wait\-Time}}

\nc\close     {\mbox{close}}
\nc\closed    {\mbox{closed}}
\nc\coming    {\mbox{coming}}
\nc\incrossing{\mbox{in\_crossing}}
\nc\open      {\mbox{open}}
\nc\opened    {\mbox{opened}}
\nc\Tracks    {\mbox{Tracks}}
\nc\voc       {\mbox{voc}}
\rnc\empty    {\mbox{empty}}

\nc\Controller{\mbox{\sc controller}} 
\nc\Gate      {\mbox{\sc gate}}  

\title{The Railroad Crossing Problem:\\
An Experiment with Instantaneous Actions and Immediate Reactions}
\author{Yuri Gurevich%
\thanks{Partially supported by NSF grant CCR-95-04375
and ONR grant N00014-94-1-1182.}
and James K. Huggins\footnotemark[1]}
\institute{EECS Department, University of Michigan, Ann Arbor,
MI, 48109-2122, USA}

\begin{document}\maketitle

\begin{abstract} 
We give an evolving algebra solution for the well-known railroad
crossing problem and use the occasion to experiment with agents that
perform instantaneous actions in continuous time and in particular
with agents that fire at the moment they are enabled.
\end{abstract}

\section{Introduction}

The well-known railroad crossing problem has been used as an example
for comparing various specification and validation methodologies; see
for example~\cite{HL,ORS} and the relevant references there.  The
evolving algebras (EA) methodology has been used extensively for
specification and validation for real-world software and hardware
systems; see the EA guide~\cite{Guide} and the EA
bibliography~\cite{Boerger}.  The merits of using ``toy'' problems as
benchmarks are debatable; not every methodology scales well to
real-world problems.  Still, toy problems are appropriate for
experimentation.  Here we present an evolving algebra solution for the
railway crossing problem and use the opportunity for experimentation
with instantaneous actions and reactions in real time.

In Sect.~2, we describe a version of the railroad crossing problem.  
It is not difficult to generalize the problem (e.g. by
relaxing our assumptions on trains) and generalize the solution
respectively.  An interested reader may view that as an exercise.

In Sect.~3, we give a brief introduction to evolving algebras (in short,
ealgebras), in order to make this paper self-contained.  We omit
many important aspects of ealgebras and refer the interested reader to a
fuller definition in the EA guide~\cite{Guide}.   In Sect.~4,
experimenting with instantaneous actions in real time, we define special
distributed real-time ealgebras appropriate to situations like that of the
railroad crossing problem.  

In Sect.~5 and Sect.~6, we give a solution for the railroad crossing
problem which is formalized as an ealgebra.  The program for the
ealgebra is given in Sect.~5.  The reader may wish to look at
Sect.~5 right away; the notation is self-explanatory to a large
extent.  In Sect.~6, we define regular runs (the only relevant runs)
of our ealgebra and analyze those runs.  Formally speaking, we have to
prove the existence of regular runs for every possible pattern of
trains; for technical reasons, we delay the existence theorem until
later.

In Sect.~7, we prove the safety and liveness properties of our
solution.  In Sect.~8 we prove a couple of additional properties of
our ealgebra.  In Sect.~9, we take advantage of the additional
properties and prove the existence theorem for regular runs and
analyze the variety of regular runs.

The ealgebra formalization is natural and this allows us to use
intuitive terms in our proofs.  One may have an impression that no
formalization is really needed.  However, a formalization is needed if
one wants a mathematical verification of an algorithm: mathematical
proofs are about mathematical objects.  Of course, we could avoid
intuitive terms and make the proofs more formal and pedantic, but this
paper is addressed to humans and it is so much harder to read pedantic
proofs.  It is a long standing tradition of applied mathematics to use
intuitive terms in proofs.  Let us notice though that more formal and
pedantic proofs have their own merits; if one wants to check the
details of our proofs by machine, it is useful to rewrite the proofs
in a pedantic way.  In any case, we see a great value in the
naturality of formalization.  No semantical approach makes inherent
difficulties of a given problem go away.  At best, the approach does
not introduce more complications and allows one to deal with the
inherent complexity of the given problem.

\paragraph{Acknowledgments.}
Raghu Mani participated in an initial stage of the work \cite{tech1}.
During the final stage of the work, the first author was
a CNRS\footnote{Centre National de la Recherche Scientifique} visitor
in the Laboratoire Informatique Theoretique et Programmation, Paris,
France \cite{tech2}.

\section{The Railroad Crossing Problem} 

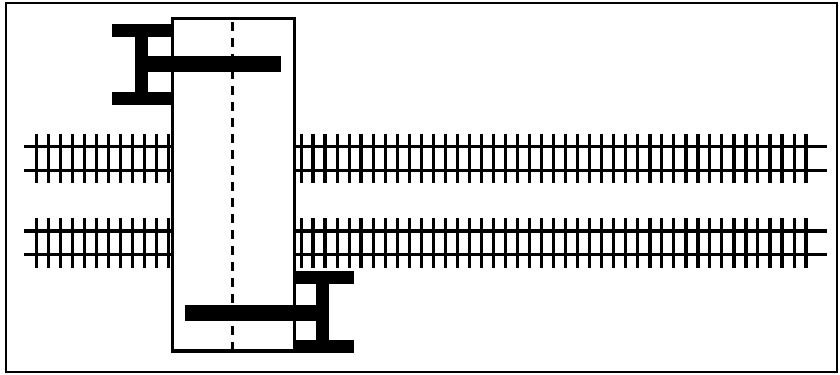
\begin{figure}[htbp]
\setlength{\unitlength}{0.00083300in}%
\begingroup\makeatletter\ifx\SetFigFont\undefined
\def\x#1#2#3#4#5#6#7\relax{\def\x{#1#2#3#4#5#6}}%
\expandafter\x\fmtname xxxxxx\relax \def\y{splain}%
\ifx\x\y   
\gdef\SetFigFont#1#2#3{%
  \ifnum #1<17\tiny\else \ifnum #1<20\small\else
  \ifnum #1<24\normalsize\else \ifnum #1<29\large\else
  \ifnum #1<34\Large\else \ifnum #1<41\LARGE\else
     \huge\fi\fi\fi\fi\fi\fi
  \csname #3\endcsname}%
\else
\gdef\SetFigFont#1#2#3{\begingroup
  \count@#1\relax \ifnum 25<\count@\count@25\fi
  \def\x{\endgroup\@setsize\SetFigFont{#2pt}}%
  \expandafter\x
    \csname \romannumeral\the\count@ pt\expandafter\endcsname
    \csname @\romannumeral\the\count@ pt\endcsname
  \csname #3\endcsname}%
\fi
\fi\endgroup
\begin{center}\framebox{
\begin{picture}(5024,2200)(66,-1417)
\thicklines
\put( 78,-63){\line( 1, 0){914}}
\put( 1764,-63){\line(1,0){3314}}
\put( 78,-213){\line( 1, 0){914}}
\put( 1764,-213){\line( 1, 0){3314}}
\put(153, 12){\line( 0,-1){300}}
\put(228, 12){\line( 0,-1){300}}
\put(303, 12){\line( 0,-1){300}}
\put(378, 12){\line( 0,-1){300}}
\put(453, 12){\line( 0,-1){300}}
\put(528, 12){\line( 0,-1){300}}
\put(603, 12){\line( 0,-1){300}}
\put(678, 12){\line( 0,-1){300}}
\put(753, 12){\line( 0,-1){300}}
\put(828, 12){\line( 0,-1){300}}
\put(903, 12){\line( 0,-1){300}}
\put(978, 12){\line( 0,-1){300}}
\put(1803, 12){\line( 0,-1){300}}
\put(1878, 12){\line( 0,-1){300}}
\put(1878, 12){\line( 0,-1){300}}
\put(2028, 12){\line( 0,-1){300}}
\put(1953, 12){\line( 0,-1){300}}
\put(2103, 12){\line( 0,-1){300}}
\put(2103, 12){\line( 0,-1){300}}
\put(2178, 12){\line( 0,-1){300}}
\put(2178, 12){\line( 0,-1){300}}
\put(2253, 12){\line( 0,-1){300}}
\put(2328, 12){\line( 0,-1){300}}
\put(2403, 12){\line( 0,-1){300}}
\put(2478, 12){\line( 0,-1){300}}
\put(2628, 12){\line( 0,-1){300}}
\put(2553, 12){\line( 0,-1){300}}
\put(2703, 12){\line( 0,-1){300}}
\put(2778, 12){\line( 0,-1){300}}
\put(2853, 12){\line( 0,-1){300}}
\put(3003, 12){\line( 0,-1){300}}
\put(2928, 12){\line( 0,-1){300}}
\put(3078, 12){\line( 0,-1){300}}
\put(3153, 12){\line( 0,-1){300}}
\put(3228, 12){\line( 0,-1){300}}
\put(3303, 12){\line( 0,-1){300}}
\put(3378, 12){\line( 0,-1){300}}
\put(3453, 12){\line( 0,-1){300}}
\put(3528, 12){\line( 0,-1){300}}
\put(3603, 12){\line( 0,-1){300}}
\put(3678, 12){\line( 0,-1){300}}
\put(3828, 12){\line( 0,-1){300}}
\put(3753, 12){\line( 0,-1){300}}
\put(3903, 12){\line( 0,-1){300}}
\put(3978, 12){\line( 0,-1){300}}
\put(4053, 12){\line( 0,-1){300}}
\put(4128, 12){\line( 0,-1){300}}
\put(4203, 12){\line( 0,-1){300}}
\put(4278, 12){\line( 0,-1){300}}
\put(4353, 12){\line( 0,-1){300}}
\put(4353, 12){\line( 0,-1){300}}
\put(4428, 12){\line( 0,-1){300}}
\put(4503, 12){\line( 0,-1){300}}
\put(4578, 12){\line( 0,-1){300}}
\put(4653, 12){\line( 0,-1){300}}
\put(4728, 12){\line( 0,-1){300}}
\put(4803, 12){\line( 0,-1){300}}
\put(4878, 12){\line( 0,-1){300}}
\put(4878, 12){\line( 0,-1){300}}
\put(4953, 12){\line( 0,-1){300}}
\put( 78,-588){\line( 1, 0){914}}
\put( 1764,-588){\line( 1, 0){3314}}
\put( 78,-738){\line( 1, 0){914}}
\put( 1764,-738){\line( 1, 0){3314}}
\put(153,-513){\line( 0,-1){300}}
\put(228,-513){\line( 0,-1){300}}
\put(303,-513){\line( 0,-1){300}}
\put(378,-513){\line( 0,-1){300}}
\put(453,-513){\line( 0,-1){300}}
\put(528,-513){\line( 0,-1){300}}
\put(603,-513){\line( 0,-1){300}}
\put(678,-513){\line( 0,-1){300}}
\put(753,-513){\line( 0,-1){300}}
\put(828,-513){\line( 0,-1){300}}
\put(903,-513){\line( 0,-1){300}}
\put(978,-513){\line( 0,-1){300}}
\put(1803,-513){\line( 0,-1){300}}
\put(1878,-513){\line( 0,-1){300}}
\put(1878,-513){\line( 0,-1){300}}
\put(2028,-513){\line( 0,-1){300}}
\put(1953,-513){\line( 0,-1){300}}
\put(2103,-513){\line( 0,-1){300}}
\put(2103,-513){\line( 0,-1){300}}
\put(2178,-513){\line( 0,-1){300}}
\put(2178,-513){\line( 0,-1){300}}
\put(2253,-513){\line( 0,-1){300}}
\put(2328,-513){\line( 0,-1){300}}
\put(2403,-513){\line( 0,-1){300}}
\put(2478,-513){\line( 0,-1){300}}
\put(2628,-513){\line( 0,-1){300}}
\put(2553,-513){\line( 0,-1){300}}
\put(2703,-513){\line( 0,-1){300}}
\put(2778,-513){\line( 0,-1){300}}
\put(2853,-513){\line( 0,-1){300}}
\put(3003,-513){\line( 0,-1){300}}
\put(2928,-513){\line( 0,-1){300}}
\put(3078,-513){\line( 0,-1){300}}
\put(3153,-513){\line( 0,-1){300}}
\put(3228,-513){\line( 0,-1){300}}
\put(3303,-513){\line( 0,-1){300}}
\put(3378,-513){\line( 0,-1){300}}
\put(3453,-513){\line( 0,-1){300}}
\put(3528,-513){\line( 0,-1){300}}
\put(3603,-513){\line( 0,-1){300}}
\put(3678,-513){\line( 0,-1){300}}
\put(3828,-513){\line( 0,-1){300}}
\put(3753,-513){\line( 0,-1){300}}
\put(3903,-513){\line( 0,-1){300}}
\put(3978,-513){\line( 0,-1){300}}
\put(4053,-513){\line( 0,-1){300}}
\put(4128,-513){\line( 0,-1){300}}
\put(4203,-513){\line( 0,-1){300}}
\put(4278,-513){\line( 0,-1){300}}
\put(4353,-513){\line( 0,-1){300}}
\put(4353,-513){\line( 0,-1){300}}
\put(4428,-513){\line( 0,-1){300}}
\put(4503,-513){\line( 0,-1){300}}
\put(4578,-513){\line( 0,-1){300}}
\put(4653,-513){\line( 0,-1){300}}
\put(4728,-513){\line( 0,-1){300}}
\put(4803,-513){\line( 0,-1){300}}
\put(4878,-513){\line( 0,-1){300}}
\put(4878,-513){\line( 0,-1){300}}
\put(4953,-513){\line( 0,-1){300}}
\put(1008,-1337){\framebox(750,2063){}}
\multiput(1377,-1337)(0,100){21}{\line( 0, 1){50}}

\linethickness{75\unitlength}
\put(630,236){\line(1,0){375}}
\put(630,665){\line(1,0){375}}
\put(810,236){\line(0,1){450}}
\linethickness{91\unitlength}
\put(845,455){\line(1,0){826}}

\linethickness{75\unitlength}
\put(1757,-1310){\line(1,0){375}}
\put(1757,-880){\line(1,0){375}}
\put(1935,-1310){\line(0,1){450}}
\linethickness{91\unitlength}
\put(1081,-1100){\line(1,0){826}}

\end{picture}}
\end{center}
\caption{\label{trainpic}A railroad crossing.}
\end{figure}

\begin{figure}[htbp]
\setlength{\unitlength}{0.00083300in}%
\begingroup\makeatletter\ifx\SetFigFont\undefined
\def\x#1#2#3#4#5#6#7\relax{\def\x{#1#2#3#4#5#6}}%
\expandafter\x\fmtname xxxxxx\relax \def\y{splain}%
\ifx\x\y   
\gdef\SetFigFont#1#2#3{%
  \ifnum #1<17\tiny\else \ifnum #1<20\small\else
  \ifnum #1<24\normalsize\else \ifnum #1<29\large\else
  \ifnum #1<34\Large\else \ifnum #1<41\LARGE\else
     \huge\fi\fi\fi\fi\fi\fi
  \csname #3\endcsname}%
\else
\gdef\SetFigFont#1#2#3{\begingroup
  \count@#1\relax \ifnum 25<\count@\count@25\fi
  \def\x{\endgroup\@setsize\SetFigFont{#2pt}}%
  \expandafter\x
    \csname \romannumeral\the\count@ pt\expandafter\endcsname
    \csname @\romannumeral\the\count@ pt\endcsname
  \csname #3\endcsname}%
\fi
\fi\endgroup
\begin{center}\framebox{
\begin{picture}(4374,1641)(64,-1198)
\thicklines
\put(1742,-1186){\framebox(847,1032){}}
\multiput(2165,-1186)(0,100){10}{\line( 0, 1){50}}
\put(1502,238){\circle{396}}
\put(1427,163){\makebox(0,0)[lb]{\smash{\SetFigFont{12}{14.4}{rm}R2}}}
\put(2852,238){\circle{396}}
\put(2777,163){\makebox(0,0)[lb]{\smash{\SetFigFont{12}{14.4}{rm}L2}}}
\put(377,238){\circle{396}}
\put(302,163){\makebox(0,0)[lb]{\smash{\SetFigFont{12}{14.4}{rm}L1}}}
\put(4127,238){\circle{396}}
\put(4052,163){\makebox(0,0)[lb]{\smash{\SetFigFont{12}{14.4}{rm}R1}}}
\put(153,-513){\line( 0,-1){300}}
\put(228,-513){\line( 0,-1){300}}
\put(303,-513){\line( 0,-1){300}}
\put(378,-513){\line( 0,-1){300}}
\put(453,-513){\line( 0,-1){300}}
\put(528,-513){\line( 0,-1){300}}
\put(603,-513){\line( 0,-1){300}}
\put(678,-513){\line( 0,-1){300}}
\put(753,-513){\line( 0,-1){300}}
\put(828,-513){\line( 0,-1){300}}
\put(903,-513){\line( 0,-1){300}}
\put(978,-513){\line( 0,-1){300}}
\put(1053,-513){\line( 0,-1){300}}
\put(1128,-513){\line( 0,-1){300}}
\put(1128,-513){\line( 0,-1){300}}
\put(1203,-513){\line( 0,-1){300}}
\put(1278,-513){\line( 0,-1){300}}
\put(1353,-513){\line( 0,-1){300}}
\put(1428,-513){\line( 0,-1){300}}
\put(1503,-513){\line( 0,-1){300}}
\put(1503,-513){\line( 0,-1){300}}
\put(1578,-513){\line( 0,-1){300}}
\put(2703,-513){\line( 0,-1){300}}
\put(2778,-513){\line( 0,-1){300}}
\put(2853,-513){\line( 0,-1){300}}
\put(3003,-513){\line( 0,-1){300}}
\put(2928,-513){\line( 0,-1){300}}
\put(3078,-513){\line( 0,-1){300}}
\put(3153,-513){\line( 0,-1){300}}
\put(3228,-513){\line( 0,-1){300}}
\put(3303,-513){\line( 0,-1){300}}
\put(3378,-513){\line( 0,-1){300}}
\put(3453,-513){\line( 0,-1){300}}
\put(3528,-513){\line( 0,-1){300}}
\put(3603,-513){\line( 0,-1){300}}
\put(3678,-513){\line( 0,-1){300}}
\put(3828,-513){\line( 0,-1){300}}
\put(3753,-513){\line( 0,-1){300}}
\put(3903,-513){\line( 0,-1){300}}
\put(3978,-513){\line( 0,-1){300}}
\put(4053,-513){\line( 0,-1){300}}
\put(4128,-513){\line( 0,-1){300}}
\put(4203,-513){\line( 0,-1){300}}
\put(4278,-513){\line( 0,-1){300}}
\put(4353,-513){\line( 0,-1){300}}
\put(4353,-513){\line( 0,-1){300}}
\put(2626,-586){\line( 1, 0){1800}}
\put( 76,-586){\line( 1, 0){1650}}
\put( 76,-736){\line( 1, 0){1650}}
\put(2626,-736){\line( 1, 0){1800}}
\put(1651,-511){\line( 0,-1){300}}
\put(301, 14){\vector(-1,-3){225}}
\put(1501, 14){\vector( 1,-3){225}}
\put(2851, 14){\vector(-1,-3){225}}
\put(4201, 14){\vector( 1,-3){225}}
\end{picture}}
\end{center}
\caption{\label{senspic}Placement of sensors along a railroad track.}
\end{figure}
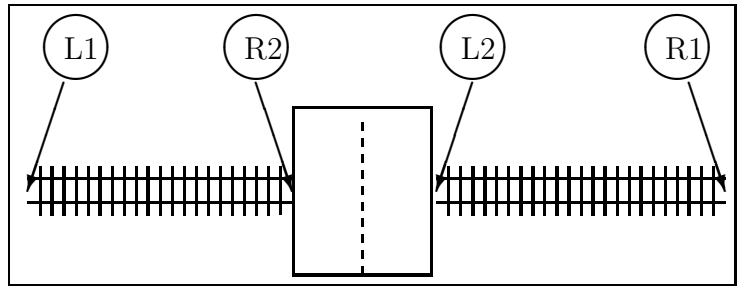

Imagine a railroad crossing with several train tracks and a common
gate, such as the one depicted in Fig.~\ref{trainpic}.  Sensors along
every track detect oncoming and departing trains.  Let us consider one
of the tracks, shown in Fig.~\ref{senspic}.  It has four sensors at
points L1, L2, R1 and R2.  Sensor L1 detects trains coming from the
left, and sensor L2 detects when those trains leave the crossing.
Similarly sensor R1 detects trains coming from the right, and sensor
R2 detects when those trains leave the crossing.  Based on signals
from these sensors, an automatic controller signals the gate to open
or close.

The problem is to design a controller that guarantees the following
requirements.
\begin{description}

\item[Safety] If a train is in the crossing, the gate is closed.

\item[Liveness] The gate is open as much as possible.
\end{description}

Several assumptions are made about the pattern of train movement.  For
example, if a train appears from the left, it leaves the crossing to
the right.  It is easiest to express those assumptions as a
restriction on possible histories of train motion on any given track.

\paragraph{Assumptions Regarding Train Motion.}
For any given track, there is a finite or infinite sequence of moments

\[ t_0 < t_1 < t_2 < t_3 < \ldots \]

\noindent
satisfying the following conditions.

\begin{description}

\item[Initial State] The moment $t_0$ is the initial moment.  The
observed part $[L1,R1]$ of the track is empty at $t_0$.

\item[Train Pattern] 
If $t_{3i+1}$ appears in the sequence then $t_{3i+3}$ appears 
in the sequence and we have that
\begin{itemize}

\item at $t_{3i+1}$, one oncoming train is detected at L1 or R1, 

\item at $t_{3i+2}$ the train reaches the crossing, and 

\item at $t_{3i+3}$ the train is detected to have left the crossing
at L2 or R2 respectively.
\end{itemize}

\item[Completeness]
There are no other trains.  

\end{description}   

\paragraph{Additional Assumptions.} From the moment that an oncoming
train is detected, it takes time between $\dmin$ and $\dmax$ for the
train to reach the crossing.  In terms of the sequence $\langle t_0 <
t_1 < t_2 < t_3 <\ldots\rangle$ above, this assumption can be stated
as follows:

\begin{description}
\item[1] Every difference $t_{3i+2} - t_{3i+1}$ belongs to the
interval $[\dmin,\dmax]$.
\end{description}

\noindent Further, the gate closes within time $\dclose$ and opens
within time $\dopen$.  This does not necessarily mean that if the
controller signals the gate to close (respectively open) at moment $t$
then the gate closes (respectively opens) by time $t +\dclose$
(respectively $t +\dopen$).  Let us state the assumption more precisely
as a restriction on possible histories.

\begin{description}
\item[2] There is no interval
$I=(t,t+\dclose)$ (respectively $I=(t,t+\dopen)$) during which the
signal to close (respectively to open) is in force but the gate is not
closed (respectively opened) at any moment in $I$.
\end{description}

\noindent It is easy to see that the controller cannot guarantee the
safety requirement is satisfied if $\dmin<\dclose$.  We ignore the
case $\dmin=\dclose$ and assume that

\begin{description}
\item[3] $\dclose <\dmin$.
\end{description}

\noindent
Finally, we will assume that actions are performed instantaneously. Of
course, real actions take time and the use of instantaneous actions is
an abstraction.  But this may be a useful abstraction.  For example,
in our case, it is natural to ignore the time taken by the
controller's actions.  It is not natural at all to view closing and
opening of the gate as instantaneous actions, and we will not do that.
Let us stress that the evolving algebra methodology does not require
that actions are necessarily instantaneous.  See for
example~\cite{BGR} where an instantaneous action ealgebra is refined
to a prolonged-action ealgebra.

The design part of the railway crossing problem  is not difficult,
especially because the problem has been addressed in a number of papers. 
What remains is to formalize the design in a specification language, in our
case as an evolving algebra, and prove the safety and liveness requirements
are satisfied.

\section{Evolving Algebras Reminder}

We give a brief reminder on evolving algebras based on the EA
guide~\cite{Guide}.  We present only what is necessary here and ignore many
important features.

\subsection{Static Algebras} 

Static algebras are essentially logicians' structures except that a
tiny bit of meta-mathematics is built into it.  They are indeed algebras in
the sense of the science of universal algebra.

A {\em vocabulary\/} is a collection of function symbols; each symbol
has a fixed arity.  Some function symbols are tagged as relation
symbols (or predicates).  It is supposed that every vocabulary
contains the following {\em logic symbols\/}: nullary symbols \true,
\false, \undef, a binary symbol =, and the symbols of the standard
propositional connectives.

A {\em static algebra\/} (or a {\em state\/}) $A$ {\em of
vocabulary\/} $\U$ is a nonempty set $X$ (the {\em basic set\/} or
{\em superuniverse\/} of $A$), together with interpretations of all
function symbols in $\U$ over $X$ (the {\em basic functions\/} of
$A$).  A function symbol $f$ of arity $r$ is interpreted as an $r$-ary
operation over $X$ (if $r=0$, it is interpreted as an element of $X$).
The interpretations of predicates ({\em basic relations\/}) and the
logic symbols satisfy some obvious requirements stated below.

Remark on notations and denotations.  A symbol in $\U$ is a name or
notation for the operation that interprets it in $A$, and the
operation is the meaning or denotation of the symbol in $A$.  In English, a
word ``spoon" is a name of a familiar table utensil, and one says ``I
like that spoon" rather than a more cumbersome ``I like that utensil
named `spoon'".  Similarly, when a state is fixed, we may say that $f$
maps a tuple $\bar{a}$ to an element $b$ rather than that the
interpretation of $f$ maps a tuple $\bar{a}$ to an element $b$.

On the interpretations of logic symbols and predicates.  Intuitively, (the
interpretations of) \true\ and \false\ represent truth and falsity
respectively.  Accordingly, the symbols \true\ and \false\ are
interpreted by different elements.  These two elements are the only
possible values of any basic relation.  The Boolean connectives behave
in the expected way over these two elements, and the equality function
behaves in the expected way over all elements.

Universes and typing.  Formally speaking, a static algebra is
one-sorted.  However, it may be convenient to view it as many-sorted;
here we describe a standard way to do this.  Some unary basic
relations are designated as universes (or sorts) and their names
may be called universe symbols.  One thinks about a universe $U$ as a
set $\{x: U(x)=\true\}$.  Basic functions are assigned universes as
domains.  For example, the domain of a binary function $f$ may be
given as $U_1\times U_2$ where $U_1$ and $U_2$ are universes.  If $f$
is a relation, this means that $f(a_1,a_2) =\false$ whenever
$a_1\not\in U_1$ or $a_2\not\in U_2$.  Otherwise this means that
$f(a_1,a_2) =\undef$ whenever $a_1\not\in U_1$ or $a_2\not\in U_2$, so
that $f$ is intuitively a partial function.

Remark on the built-in piece of meta-mathematics.  In first-order
logic, an assertion about a given structure does not evaluate to any
element of the structure.  For technical convenience, in evolving
algebras truth and falsity are represented internally and many
assertions can be treated as terms.  This technical modification does
not prevent us from dealing with assertions directly.  For example,
let $f,g$ be nullary function symbols and $P$ a binary function
symbol.  Instead of saying that $P(f,g)$ evaluates to \true\
(respectively \false) at a state $A$, we may say $P(f,g)$ holds
(respectively fails) at $A$.  In some cases, we may even omit ``holds'';
for example, we may assert simply that $f\neq g$.  Admittedly, this
is not very pedantic, but we write for humans, not machines.

\subsection{Updates}

Alternatively, a state can be viewed as a kind of memory.
A {\em location\/} $\ell$ of a state $A$ of vocabulary $\U$ is a pair
$\ell = (f,\bar{a})$ where $f$ is a symbol in $\U$ of some arity $r$
and $\bar{a}$ is an $r$-tuple of elements of $A$ (that is, of the
superuniverse of $A$).  The element $f(\bar{a})$ is the {\em content\/}
of location $\ell$ in $A$.

An {\em update} of state $A$ is a pair $(\ell,b)$, where $\ell$ is some
location $(f,\bar{a})$ of $A$ and $b$ is an element of $A$; it is
supposed that $b$ is (the interpretation of) \true\ or \false\ if $f$
is a predicate.   This update
is {\em trivial\/} if $b$ is the content of $\ell$ in $A$.  An update can be
performed: just replace the value at location $\ell$ with $b$.  The
vocabulary, the superuniverse and the contents of other locations remain
unchanged.  The state changes only if the update is nontrivial.

Call a set $S =\{(\ell_1,b_1),\ldots,(\ell_n,b_n)\}$ of updates of a state $A$
{\em consistent\/} if the locations are distinct.  In other words,  $S$ is {\em
inconsistent\/} if there are $i,j$ such that $\ell_i =\ell_j$ but $b_i\neq b_j$. 
In the case that $S$ is consistent it is performed as follows:  replace the
content of $\ell_1$ with $b_1$, the content of $\ell_2$ with $b_2$  and so on. 
To perform an inconsistent update set, do nothing.

A pedantic remark.  The equality used in the previous paragraph is
not the built-in equality of $A$ but rather the equality of the meta
language.  One could use another symbol for the built-in equality, but
this is not necessary.   

A remark to theoreticians.  At the point that updates are introduced,
some people, in particular Robin Milner~\cite{Milner}, raise an
objection that an update may destroy algebraic properties.  For
example, an operation may lose associativity.
That is true.  So, in what sense are
static algebras algebraic?  They are algebraic in the sense that the
nature of elements does not matter and one does not distinguish
between isomorphic algebras.  A standard way to access a particular
element is to write a term that evaluates to that element.  Coming
back to algebraic properties like associativity (and going beyond the
scope of this paper), let us note that, when necessary, one can
guarantee that such a property survives updating by declaring
some functions static or by imposing appropriate integrity constraints
or just by careful programming.

\subsection{Basic Rules} 

In this subsection we present the syntax and semantics of basic rules. 
Each rule $R$ has a vocabulary, namely the collection of function symbols
that occur in $R$.  A rule $R$ is applicable to a state $A$ only if the
vocabulary of $A$ includes that of $R$.  At each state $A$ of sufficiently
rich vocabulary, $R$ gives rise to a set of updates.  To execute $R$ at 
such a state $A$, perform the update set at $A$. 

A {\em basic update rule\/} $R$ has the form

\begin{tabbing}mmmmm\=\kill
\> $f(e_1,\ldots,e_r) := e_0$\
\end{tabbing}

\noindent
where $f$ is an $r$-ary function symbol (the {\em head\/} of $R$) and
each $e_i$ is a ground term, that is, a term without any variables.
(In programming languages, terms are usually called expressions; that
motivates the use of letter e for terms.)  To execute $R$ at a state
$A$ of sufficiently rich vocabulary, evaluate all terms $e_i$ at $A$
and then change $f$ accordingly.  In other words, the update set
generated by $R$ at $A$ consists of one update $(\ell,a_0)$ where
$\ell = (f,(a_1,\ldots,a_r))$ and each $a_i$ is the value of $e_i$ at
$A$.

For example, consider an update rule $f(c_1 + c_2) := c_0$ and a state
$A$ where $+$ is interpreted as the standard addition function on
natural numbers and where $c_1,c_2,c_0$ have values $3, 5, 7$
respectively.  To execute the rule at $A$, set $f(8)$ to $7$.

There are only two basic rule constructors.  
One is the {\em conditional constructor\/} which
produces rules of the form: 

\begin{tabbing}mmmmm\=\kill
\>{\bf if} $g$ {\bf then} $R_1$ {\bf else} $R_2$ {\bf endif}
\end{tabbing}

\noindent where $g$ is a ground term (the {\em guard\/} of the new rule) and
$R_1,R_2$ are rules.  To execute the new rule in a state $A$ of
sufficiently rich vocabulary, evaluate the guard.  If it is true, then
execute $R_1$; otherwise execute $R_2$.  (The ``{\bf else}'' clause
may be omitted if desired.)

The other constructor is the {\em block constructor\/} which produces
rules of the form:

\begin{tabbing}mmmmm\=mmmmm\=\kill
\>{\bf block}\\
\> \> $R_1$\\
\> \> $\vdots$\\
\> \> $R_k$\\
\>{\bf endblock}
\end{tabbing}

\noindent where $R_1,\ldots,R_k$ are rules.  (We often omit the
keywords ``{\bf block}'' and ``{\bf endblock}'' for brevity and use
indentation to eliminate ambiguity.)  To execute the new rule in a
state $A$ of sufficiently rich vocabulary, execute rules
$R_1,\ldots,R_k$ simultaneously.  More precisely, the update set
generated by the new rule at $A$ is the union of the update sets
generated by the rules $R_i$ at $A$.

A {\em basic program\/} is simply a basic rule.  

In this paper we say that a rule $R$ is {\em enabled\/} at a state $A$ of
sufficiently rich vocabulary if the update set generated by $R$ at $A$ is
consistent and contains a non-trivial update; otherwise $R$ is {\em
disabled\/} at $A$.  (The notion of being enabled has not been formalized in
the EA guide.)  Rules will be executed only if they are enabled, so that
the execution changes a given state.  This seems to be a very pedantic
point.  What harm is done by executing a rule that does not change a given
state?  It turns out that the stricter notion of being enabled is
convenient in real-time computational theory; see Lemma~\ref{lem4.4} in this
connection.   

\subsection{Parallel Synchronous Rules} 

Generalize the previous framework in two directions.  First, permit
terms with variables and generalize the notion of state: in addition
to interpreting some function names, a generalized state may assign
values to some variables.  (Notice that a variable cannot be the head
of an update rule.)

Second, generalize the notion of guards by allowing bounded
quantification.  More formally, we define {\em guards\/} as a new
syntactical category.  Every term $P(e_1,\ldots,e_r)$, where $P$
is a predicate, is a guard.  A Boolean combination of guards is a
guard.  If $g(x)$ is a guard with a variable $x$ and $U$ is a universe
symbol then the expression $(\forall x\in U) g(x)$ is also a guard.

The semantics of guards is quite obvious.  A guard $g(\by)$ with free
variables $\by$ holds or fails at a (generalized) state $A$ that
assigns values to all free variables of $g$.  The least trivial case
is that of a guard $g(\by) = (\forall x\in U) g'(x,\by)$.  For every
element $b$ of $U$ in $A$, let $A_b$ be the expansion of $A$ obtained
by assigning the value $b$ to $x$.  Then $g(\by)$ holds at $A$ if
$g'(x,\by)$ holds at every $A_b$; otherwise it fails at $A$.

Now consider a generalized basic rule $R(x)$ with a variable $x$ and
let $U$ be a universe symbol.  Form the following rule $R^*$:

\begin{tabbing}mmmmm\=mmmmm\=\kill
\>{\bf var} $x$ {\bf ranges over} $U$\\
\> \>$R(x)$\\
\>{\bf endvar}
\end{tabbing}

Intuitively, to execute $R^*$, one executes $R(x)$ for every $x\in U$.
To make this more precise, let $A$ be a (generalized) state that
interprets all function names in the vocabulary of $R(x)$ and assigns
values to all free variables of $R(x)$ except for $x$.  For each
element $b$ of the universe $U$ in $A$, let $A_b$ be the expansion of
$A$ obtained by assigning the value $b$ to $x$, and let $E_b$ be the
update set generated by $R(x)$ at $A_b$.  Since $x$ does not appear as
the head of any update instruction in $R(x)$, each $E_b$ is also a set
of updates of $A$.  The update set generated by $R^*$ at $A$ is the
union of the update sets $E_b$.

Call the new rule a {\em parallel synchronous rule\/} (or a {\em
declaration rule\/}, as in the EA guide).  A {\em parallel synchronous
program\/} is simply a parallel synchronous rule without free variables.
Every occurrence of a variable should be bound by a declaration or a
quantifier.

\subsection{Special Distributed Programs}

For our purposes here, a {\em distributed program\/} $\Pi$ is given by a
vocabulary and a finite set of basic or parallel synchronous programs
with function symbols from the vocabulary of $\Pi$.  The constitutent
programs are the {\em modules\/} of $\cA$.  A {\em state\/} of $\Pi$ is a
state of the vocabulary of $\Pi$.  
Intuitively, each module is executed by a separate agent.

This is a very restricted definition.  For example, the EA guide allows
the creation of new agents during the evolution.  

Intuitively, it is convenient though to distinguish between a module (a
piece of syntax) and its executor, and even think about agents in
anthropomorphic terms.  But since in this case agents are uniquely defined by
their programs, there is no real need to have agents at all, and we may
identify an agent by the name of its program.

\section{Special Distributed Real-Time Ealgebras}

A program does not specify a (distributed) ealgebra completely. 
We need to define what constitutes a computation (or a run) and
then to indicate initial states and maybe a relevant class of runs.
In this section, we define a restricted class of distributed
real-time evolving algebras by restricting attention to static algebras of
a particular kind and defining a particular notion of run.  

We are interested in computations in real time that satisfiy the following
assumptions.

\begin{description}

\item[I1] Agents execute instantaneously.

\item[I2] Enviromental changes take place instantaneously.

\item[I3] The global state of the given distributed ealgebra is well defined
at every moment.

\end{description}

Let us stress again that the three assumptions above are not a part of the
evolving algebra definition.  The prolonged-action ealgebra~\cite{BGR},
mentioned in Sect.~2, satisfies none of these three assumptions.

\paragraph{Vocabularies and Static Structures.}

Fix some vocabulary $\U$ with a universe symbol Reals and let $\U^+$ be the
extension of $\U$ with a nullary function symbol CT; it is supposed of
course that $\U$ does not contain CT.  Restrict attention to $\U^+$-states
where the universe Reals is the set of real numbers and CT evaluates to a
real number.  Intuitively, CT gives the current time.  

\subsection{Pre-runs}

\begin{definition}\label{Definition of pre-runs}
A {\em pre-run\/} $R$ of vocabulary $\U^+$ is a mapping from the interval
$[0,\infty)$ or the real line to states of vocabulary $\U^+$ satisfying the
following requirements where $\r(t)$ is the reduct of $R(t)$ to $\U$.
\begin{description}

\item[Superuniverse Invariability] 
The superuniverse does not change during the evolution; that is, the
superuniverse of every $R(t)$ is that of $R(0)$. 

\item[Current Time]
At every $R(t)$, $\CT$ evaluates to $t$.

\item[Discreteness] 
For every $\t>0$,  there is a finite sequence  $0 = t_0 < t_1 <\ldots < t_n
=\t$ such that  if $t_i < \a < \b <t_{i+1}$ then $\r(\a) =\r(\b)$.  \qed

\end{description}
\end{definition}

Remarks. Of course, we could start with an initial moment different from
$0$, but without loss of generality we can assume that the initial moment is
$0$.  Our discreteness requirement is rather simplistic (but sufficient for
our purposes in this paper).  One may have continuous time-dependent basic
functions around (in addition to CT); in such cases, the discreteness
requirement becomes more subtle.  

In the rest of this section, $R$ is a pre-run of vocabulary  $\U^+$ and
$\r(t)$ is the reduct of $R(t)$ to $\U$.

The notation $\r(t+)$ and $\r(t-)$ is self-explanatory; still, let
us define it precisely.  $\r(t+)$ is any state $\r(t+\e)$ such that
$\e>0$ and $\r(t+\d) = \r(t+\e)$ for all positive $\d<\e$.
Similarly, if $t>0$ then $\r(t-)$ is any state $\r(t-\e)$ such that
$0<\e\leq t$ and $\r(t-\d) = \r(t-\e)$ for all positive $\d<\e$.

Call a moment $t$ {\em significant\/} for $R$ if (i)~$t=0$
or (ii)~$t>0$ and either $\r(t)\neq\r(t-)$ or $\r(t)\neq\r(t+)$. 

\begin{lemma}\label{Lemma 4.1}
For any moment $t$, $\r(t+)$ is well defined.  For any moment $t>0$,
$\r(t-)$ is well defined.    If there are infinitely many significant
moments then their supremum equals $\infty$.
\end{lemma}

\begin{proof}
Obvious.
\qed\end{proof}

Recall that a set $S$ of nonnegative reals is {\em discrete\/} if it has no
limit points.  In other words, $S$ is discrete if and only if, for every
nonnegative real $\t$, the set $\{t\in S: t <\t\}$ is finite.   The
discreteness requirement in the definition of pre-runs means exactly that
the collection of the significant points of $R$ is discrete. 

We finish this subsection with a number of essentially self-evident
definitions related to a given pre-run $R$.  Let $e$ be a term of
vocabulary $\U^+$.  If $e$ has free variables then fix the values of
those variables, so that $e$ evaluates to a definite value in every
state of vocabulary $\U^+$.  (Formally speaking $e$ is a pair of the
form $(e',\xi)$ where $e'$ is a term and $\xi$ assigns elements of
$R(0)$ to free variables of $e'$.)

{\em The value $e_t$ of $e$ at moment $t$\/} is the value of $e$ in $R(t)$. 
Accordingly, $e$ {\em holds\/} (respectively {\em fails\/}) at $t$ if it does
so in $R(t)$.  Likewise, a module is {\em enabled\/} (respectively {\em
disabled\/}) at $t$ if it is so in $R(t)$.  In a similar vein, we speak about
a time interval $I$.  For example, $e$ {\em holds over\/} $I$ if it holds at
every $t\in I$.

If $e$ has the same value over some nonempty interval $(t,t+\e)$, then this
value is {\em the value $e_{t+}$ of $e$ at $t+$\/} (respectively {\em at
$t-$\/}).  Similarly, if $t>0$ and $e$ has the same value over some nonempty
interval $(t-\e,t)$, then this value is {\em the value $e_{t-}$ of $e$ at
$t-$\/}.  Define accordingly when $e$ holds, fails at $t+, t-$ and when an
agent is enabled, disabled at $t+, t-$. 

Further, $e$ {\em is set to\/} a value $a$ (or simply {\em becomes\/} $a$)
at $t$ if either (i)~$e_{t-}\neq a$ and $e_t = a$, 
or else (ii)~$e_t\neq a$ and $e_{t+}=a$.
Define accordingly when an agent becomes enabled, disabled at $t$.

\subsection{Runs}

Now consider a distributed program $\Pi$ with function symbols from
vocabulary $\U^+$.  Runs of $\Pi$ are pre-runs with some restrictions
on how the basic functions evolve.  Depending upon their use, the
basic functions of $\Pi$ fall into the 
following three disjoint categories.
\begin{description}

\item[Static] These functions do not change during any run.
The names of these functions do not appear as the heads of update
rules in $\Pi$. 

\item[Internal Dynamic] These 
functions may be changed only by agents.
The names of these functions appear as the heads of update rules and
the functions are changed by executing the modules of $\Pi$.
For brevity, we abbreviate ``internal dynamic'' to ``internal''.

\item[External Dynamic] These functions may be changed only by the
environment.  The names of these functions do not appear as the heads
of update rules; nevertheless the functions can change from 
one state to another.  Who changes them?  The environment. 
Some
restrictions may be imposed on how these functions can change.
For brevity, we abbreviate ``external dynamic'' to ``external''. 

\end{description}

Remark.  It may be convenient to have functions that can by changed
both by agents and the environment.  The EA guide allows that, but we
do not need that generality here.

Before we give the definition of runs, let us explain informally that
one should be cautious with instantaneous actions.  In particular, it
may not be possible to assume that agents always fire at the moment
they become enabled.  Consider the following two interactive
scenarios.

\begin{description}
\item[Scenario 1] The environment changes a nullary external function
$f$ at moment $t$.  This new value of $f$ enables an agent $X$.  The
agent fires immediately and changes another nullary function $g$.
\end{description}

What are the values of $f$ and $g$ at time $t$, and at what time does
$X$ fire?  If $f$ has its old value at $t$ then $X$ is disabled at $t$
and fires at some time after $t$; thus $X$ does not fire immediately.
If $g$ has its new value already at $t$ then $X$ had to fire at some
time before $t$; that firing could not be triggered by the change of
$f$.  We arrive at the following conclusions: $f$ has its new value at
$t$ (and thus $f_t$ differs from $f_{t-}$), $X$
fires at $t$, and $g$ has its old value at $t$ (and thus 
$g_t$ differs from $g_{t+}$).

\begin{description}
\item[Scenario 2] At time $t$, an agent $X$ changes a function $g$ and
in so doing enables another agent $Y$ while disabling himself.
\end{description}

When does $Y$ fire?  Since $X$ fires at $t$, it is enabled at $t$ and
thus $g$ has its old value at $t$.  Hence $Y$ is disabled at $t$ and
fires at some time after $t$.  Thus $Y$ cannot react immediately.

The following definition is designed to allow immediate agents.

\begin{definition}\label{Definition of runs} 
A pre-run $R$ of vocabulary $\U^+$ is a {\em run\/} of $\Pi$ if it
satisfies the following conditions where $\r(t)$ is the reduct of $R$
to $\U$.
\begin{enumerate}

\item 
If $\r(t+)$ differs from $\r(t)$ then $\r(t+)$ is the $\U$-reduct of
the state resulting from executing some modules $M_1,\ldots,M_k$
at $R(t)$.  In
such a case we say $t$ is {\em internally significant\/}
and the executors of $M_1,\ldots,M_k$ {\em fire\/} at $t$.  All
external functions with names in $\U$ 
have the same values in $\r(t)$ and $\r(t+)$.  

\item
If $i > 0$ and $\r(\t)$ differs from $\r(\t-)$ then they differ only
in the values of external functions.  In such a case we say
$\t$ is {\em externally significant\/}.  All internal functions
have the same values in $\r(t-)$ and $\r(t)$.    \qed
\end{enumerate}
\end{definition}

Remark.  Notice the global character of the definition of firing.  An agent
fires at a moment $t$ if $\r(t+)\neq\r(t)$.  This somewhat simplified
definition of firing is sufficient for our purposes in this paper.

In the rest of this section, $R$ is a run of $\Pi$ and $\r(t)$ the reduct
of $R(t)$ to $\U$.  Let $e$ be a term $e$ with fixed values of all its free
variables.  A
moment $t$ is {\em significant for $e$\/} if, for every $\e>0$, there exists a
moment $\a$ such that $|\a - t| <\e$ and $e_a\neq e_t$.  Call $e$ {\em
discrete\/} (in the given run $R$) if the collection of significant moments
of $e$ is discrete.  In other words, $e$ is discrete if and only, for every
$t>0$, there is a finite sequence

\[ 0 = t_0 < t_1 < \ldots < t_n = t\]

\noindent 
such that if $t_i <\a <\b < t_{i+1}$ then $e_\a = e_\b$. 

\begin{lemma}[(Discrete Term Lemma)]
If a term $e$ is discrete then 
\begin{enumerate}

\item   For every $t$, $e$ has a value at $t+$.

\item   For every $t>0$, $e$ has a value at $t-$.

\end{enumerate}
\end{lemma}

\begin{proof}
Obvious.
\qed\end{proof}  

\begin{lemma}[(Preservation Lemma)]
Suppose that a term $e$ with fixed values of its free variables does not
contain CT.  Then $e$ is discrete.  Furthermore,
\begin{enumerate}

\item  If $e$ contains no external functions and $t>0$ then $e_t = e_{t-}$.

\item  If $e$ contains no internal functions then $e_{t+} = e_t$.

\end{enumerate}
\end{lemma}

\begin{proof} 
This is an obvious consequence of the definition of runs.
\qed\end{proof}

It may be natural to have agents that fire the instant they are
enabled.

\begin{definition}
An agent is {\em immediate\/} if it fires at every state where it is
enabled.
\qed\end{definition}

\begin{lemma}[(Immediate Agent Lemma)]\label{lem4.4}\ 
\begin{enumerate}

\item  The set of moments when an immediate agent is enabled is discrete.  

\item  If the agent is enabled at some moment $t$ then it is disabled at
$t+$ and, if $t>0$, at $t-$.
\end{enumerate}
\end{lemma}

\begin{proof}\ 
\begin{enumerate}
\item
If the agent is enabled at a moment $t$, it fires at $t$ and therefore
(according to our notion of being enabled) changes the state; it follows
that $t$ is a significant moment of the run.  By the discreteness condition
on pre-runs, the collection of significant moments of a run is discrete. 
It remains to notice that every subset of a discrete set is discrete.   

\item
Follows from 1. 
\qed
\end{enumerate}
\end{proof}

Recall the scenario S2.  There agent $Y$ cannot be immediate.
Nevertheless, it may make sense to require that some agents cannot
delay firing forever.

\begin{definition}\label{Definition of bounded agent}
An agent X is {\em bounded\/} if it is immediate or there exists a 
bound
$b>0$ such that there is no interval $(t,t+b)$ during which $X$ is
continuously enabled but does not fire.
\qed\end{definition}

Notice that it is not required that if a bounded agent $X$ becomes
enabled at some moment $\a$, then it fires at some moment $\b < \a +
b$.  It is possible a priori that X becomes disabled and does not fire in that
interval.

\section{The Ealgebra for Railroad Crossing Problem}

We present our solution for the railroad crossing problem formalized
as an evolving algebra $\cA$ of a vocabulary $\U^+ =\U\cup\{\CT\}$.
In this section, we describe the program and initial states of $\cA$;
this will describe the vocabulary as well.  The relevant runs of $\cA$
will be described in the next section.

The program of $\cal A$ has two modules \Gate\ and \Controller,
shown in Fig.~\ref{rules}.

\begin{figure}[htbp]
\begin{center}
\begin{minipage}{3.5in}
\begin{tabbing}mmm\=mmm\=mmm\=mmm\=mmm\=\kill
\Gate\\
\>{\bf if} Dir = open {\bf then} GateStatus := open {\bf endif}\\
\>{\bf if} Dir = close {\bf then} GateStatus := closed {\bf endif}\\
\\
\Controller\\
\>{\bf var} $x$ {\bf ranges over} Tracks\\
\>\>{\bf if} TrackStatus$(x)$ = coming {\bf and} Deadline$(x) = \infty$ {\bf then}\\
\>\>\>Deadline$(x)$ := \CT + WaitTime\\
\>\>{\bf endif}\\
\>\>{\bf if} $\CT = $Deadline$(x)$ {\bf then} Dir := close {\bf endif}\\
\>\>{\bf if} TrackStatus$(x)$ = empty {\bf and} Deadline$(x)< \infty$ {\bf then}\\
\>\>\>Deadline$(x)$ := $\infty$\\
\>\>{\bf endif}\\
\>{\bf endvar}\\
\>{\bf if} Dir=close {\bf and} SafeToOpen {\bf then} Dir := open
{\bf endif}\\
\end{tabbing}
\end{minipage}
\end{center}
\caption{\label{rules}Rules for \Gate\ and \Controller.}
\end{figure}

Here \WT\ abbreviates the term $\dmin - \dclose$, and SafeToOpen
abbreviates the term

\[(\forall x\in\Tracks) 
[\TS(x) = \empty \mbox{\ \ or\ \ } \CT + \dopen < \DL(x)].\]

We will refer to the two constituent rules of \Gate\ as OpenGate,
CloseGate respectively.  We will refer to the three constituent rules
of \Controller's parallel synchronous rule as $\SD(x)$, $\SC(x)$,
$\CD(x)$, respectively, and the remaining conditional rule as $\SO$.

Our \GS\ has only two values: opened and closed.  This is of course a
simplification.  The position of a real gate could be anywhere between
fully closed and fully opened.  (In \cite{HL}, the position of the
gate ranges between $0^o$ and $90^o$.)  But this simplification is
meaningful.  The problem is posed on a level of abstraction where it
does not matter whether the gate swings, slides, snaps or does
something else; it is even possible that there is no physical gate,
just traffic lights.  Furthermore, suppose that the gate is opening
and consider its position as it swings from $0^o$ to $90^o$.  Is it
still closed or already open at $75^o$?  One may say that it is
neither, that it is opening.  But for the waiting cars, it is still
closed.  Accordingly \GS\ is intended to be equal to closed at this
moment.  It may change to opened when the gate reaches $90^o$.
Alternatively, in the case when the crossing is equipped with traffic
lights, it may change to opened when the light becomes green.
Similarly, it may change from opened to closed when the light becomes
red.  If one is interested in specifying the gate in greater detail,
our ealgebra can be refined by means of another ealgebra.

The program does not define our evolving algebra $\cA$ completely.  In
addition, we need to specify a collection of {\em initial states\/} and
relevant runs.

Initial states of $\cA$ satisfy the following conditions:
\begin{enumerate}
\item
The universe Tracks is finite.  The universe ExtendedReals
is an extension of the universe Reals with an additional element
$\infty$.  The binary relation $<$ and the binary operation $+$ are
standard; in particular $\infty$ is the largest element of
ExtendedReals.

\item The nullary functions close and open are interpreted by different
elements of the universe Directions.  The nullary functions closed and
opened are interpreted by different elements of the universe
GateStatuses.  The nullary functions empty, coming, \incrossing\ are
different elements of the universe TrackStatuses.

\item The nullary functions
$\dclose, \dopen, \dmax, \dmin$ are positive reals such that
\[ \dclose < \dmin \leq \dmax . \]
One may assume for simplicity of understanding that these four
reals are predefined: that is, they have the same value in all initial
state.  This assumption is not necessary.

\item The unary function \TS\ assigns (the element called) empty to
every track (that is, to every element of the universe Tracks).  The unary
function Deadline assigns $\infty$ to every track.
\end{enumerate}
It is easy to see that, in any run, every value of the internal function \DL\
belongs to ExtendedReals.  

\section{Regular Runs}

The following definition takes into account the assumptions of
Sect.~2.

\subsection{Definitions} 

\begin{definition} 
A run $R$ of our evolving algebra is {\em regular\/} if it satisfies
the following three conditions.
\begin{description}

\item[Train Motion] For any track $x$, there is a finite or infinite
sequence 
\[ 0=t_0 < t_1 < t_2 < t_3 < \ldots \] 
\noindent of so-called {\em significant moments of track $x$\/} such that
\begin{itemize}
\item
$\TS(x)$ = empty holds over every interval $[t_{3i}, t_{3i+1})$;

\item
$\TS(x)$ = coming holds over every interval $[t_{3i+1},t_{3i+2})$,
and\\ $\dmin \leq (t_{3i+2} - t_{3i+1}) \leq \dmax$;

\item 
$\TS(x) = \incrossing$ holds over every interval $[t_{3i+2},
t_{3i+3})$; and

\item if $t_k$ is the final significant moment in the sequence, then
$k$ is divisible by $3$ and $\TS(x)=\empty$ over $[t_k,\infty)$.
\end{itemize}

\item[Controller Timing] Agent \Controller\ is immediate.

\item[Gate Timing] Agent \Gate\ is bounded.  Moreover, there is no
time interval $I=(t,t+\dclose)$ such that [Dir=close and $\GS
=\opened$] holds over $I$.  Similarly there is no interval
$I=(t,t+\dopen)$ such that [Dir=open and $\GS =\closed$] holds over
$I$.  \qed
\end{description} 
\end{definition}

In the rest of this paper, we restrict attention to regular runs of
$\cA$.  Let $R$ be a regular run and $\r$ be the reduct of $R$ to
$\U$.

\subsection{Single Track Analysis}

Fix a track $x$ and let $0=t_0<t_1<t_2<\ldots$ be the significant moments
of $x$.

\begin{lemma}[(Deadline Lemma)]\ 
\begin{enumerate}

\item $\DL(x) =\infty$ over $(t_{3i},t_{3i+1}]$, and $\DL(x) = t_{3i+1} +\WT$
over $(t_{3i+1},t_{3i+3}]$.

\item Let $\Dclose = \dclose + (\dmax - \dmin) = \dmax - \WT$.  
If $\TS(x)\neq\incrossing$ over an interval $(\a,\b)$, then
Dead\-line$(x)\geq\b -\Dclose$ over $(\a,\b)$. 

\end{enumerate}
\end{lemma}

\begin{proof}\ 
\begin{enumerate}
\item
A quite obvious induction along the sequence

\[ (t_0,t_1], (t_1,t_3], (t_3,t_4], (t_4,t_6], \ldots. \]

\noindent
The basis of induction.  We prove that $\DL(x) =\infty$ over $I =
(t_0,t_1)$; it will follow by Preservation Lemma that $\DL(x) =\infty$
at $t_1$.  Initially, $\DL(x) =\infty$.  Only $\SD(x)$ can alter that
value of $\DL(x)$, but $\SD(x)$ is disabled over $(t_0,t_1)$. 
The induction step splits into two cases.

\paragraph{Case 1.}  Given that $\DL(x) =\infty$ at $t_{3i+1}$, we
prove that $\DL(x) = t_{3i+1} +\WT$ over $I = (t_{3i+1},t_{3i+3})$; it
will follow by Preservation Lemma that $\DL(x) = t_{3i+1} +\WT$ at
$t_{3i+3}$.  $\SD(x)$ is enabled and therefore fires at $t_{3i+1}$
setting $\DL(x)$ to $t_{3i+1} +\WT$.  $\CD(x)$ is the only rule that
can alter that value of $\DL(x)$ but it is disabled over $I$ because
$\TS(x)\neq\empty$ over $I$.

\paragraph{Case 2.}  Given that $\DL(x) <\infty$ at $t_{3i}$ where
$i>0$, we prove that $\DL(x) =\infty$ over $I = (t_{3i},t_{3i+1})$; it
will follow by Preservation Lemma that $\DL(x) =\infty$ at $t_{3i+1}$.
$\CD(x)$ is enabled and therefore fires at $t_{3i}$ setting $\DL(x)$
to $\infty$.  Only $\SD(x)$ can alter that value of $\DL(x)$ but it is
disabled over $I$ because $\TS(x) =\empty\neq\coming$ over $I$.

\item By contradiction suppose that $\DL(x) <\b -\Dclose$ at some
$t\in(\a, \b)$.  By 1, there is an $i$ such that $t_{3i+1}< t\leq
t_{3i+3}$ and $\DL(x) = t_{3i+1} +\WT$ at $t$.  Since $(\a,\b)$ and
the \incrossing\ interval $[t_{3i+2},t_{3i+3})$ are disjoint, we have
that $t_{3i+1} < t < \b\leq t_{3i+2}$. By the definition of regular
runs, $\dmax\geq t_{3i+2} - t_{3i+1}\geq\b - t_{3i+1}$, so that
$t_{3i+1}\geq\b -\dmax$.  We have
\[\begin{array}{rclcl}
\b-\Dclose & > & \DL(x) \mbox{ at t } & = & t_{3i+1}+\WT \\
 & \geq & \b-\dmax + \WT & = & \b-\Dclose
\end{array}\]
\noindent which is impossible. \qed
\end{enumerate}
\end{proof}

\begin{corollary}[(Three Rules Corollary)]\ 
\begin{enumerate}

\item $\SD(x)$ fires exactly at moments $t_{3i+1}$, that is exactly when
Track\-Status$(x)$ becomes coming. 

\item $\SC(x)$ fires exactly at moments $t_{3i+1} +\WT$.

\item $\CD(x)$ fires exactly at moments $t_{3i}$ with
$i>0$, that is exactly when $\TS(x)$ becomes empty.

\end{enumerate}
\end{corollary}

\begin{proof}
Obvious.
\qed\end{proof}

Let $s(x)$ be the quantifier-free part
\[ \TS(x) = \empty \mbox{\ \ or\ \ } \CT + \dopen < \DL(x). \]
of the term SafeToOpen with the fixed value of $x$.

\begin{lemma}[(Local SafeToOpen Lemma)]\ 
\begin{enumerate}

\item Suppose that $\WT >\dopen$.  Then $s(x)$ holds over intervals
$[t_{3i},t_{3i+1} +\WT -\dopen)$ (the {\em maximal positive intervals of
$s(x)$\/}) and fails over intervals $[t_{3i+1} +\WT -\dopen, t_{3i+3})$.

\item Suppose that $\WT\leq\dopen$.  Then $s(x)$ holds over intervals
$[t_{3i},t_{3i+1}]$ (the {\em maximal positive intervals of
$s(x)$\/}) and fails over intervals $(t_{3i+1}, t_{3i+3})$.

\item The term $s(v)$ is discrete.

\item $s(x)$ becomes true exactly at moments $t_{3i}$ with $i>0$, that is
exactly when $\TS(x)$ becomes empty.

\item If $[\a,\b)$ or $[\a,\b]$ is a maximal positive interval of
$s(x)$, then $\SC(x)$ is disabled over $[\a,\b]$ and at $\b+$.

\end{enumerate}
\end{lemma}

\begin{proof}\ 
\begin{enumerate}
\item
Over $[t_{3i},t_{3i+1})$, $\TS(x) =\empty$ and
therefore $s(x)$ holds.  At $t_{3i+1}$, $\DL(x) =\infty$ and therefore
$s(x)$ holds.  $\SD(x)$ fires at $t_{3i+1}$ and sets $\DL(x)$ to
$t_{3i+1} +\WT$.  Over $(t_{3i},t_{3i+1} +\WT -\dopen)$, 
\begin{eqnarray*}
\CT+\dopen &<& (t_{3i+1}+\WT-\dopen)+\dopen\\
 &=& t_{3i+1}+\WT = \DL(x)
\end{eqnarray*}
\noindent
and therefore $s(x)$ holds.  Over the interval $[t_{3i+1} +\WT
-\dopen, t_{3i+3})$, $\TS(x)\neq\empty$ and $\CT +\dopen\geq t_{3i+1}
+\WT =\DL(x)$ and therefore $s(x)$ fails.

\item The proof is similar to that of 1.

\item This follows from 1 and 2.

\item This follows from 1 and 2.

\item We consider the case when $\WT >\dopen$; the case when
$\WT\leq\dopen$ is similar.   By 1, the maximal open interval of $s(x)$
has the form $[\a,\b) = [t_{3i},t_{3i+1} +\WT -\dopen)$ for some $i$.  By
Three Rules Corollary, $\SC(x)$ fires at moments $t_{3j+1} +\WT$.  Now the
claim is obvious.  
\qed
\end{enumerate}
\end{proof}

\subsection{Multiple Track Analysis}

\begin{lemma}[(Global SafeToOpen Lemma)]\ 
\begin{enumerate}

\item The term SafeToOpen is discrete.

\item If SafeToOpen holds at $t+$ then it holds at $t$.

\item If SafeToOpen becomes true at $t$ then some $\TS(x)$ becomes empty at
$t$.

\item If SafeToOpen holds at $t$ then $t$ belongs to an interval
$[\a,\b)$ (a {\em maximal positive interval of\/} SafeToOpen) such that
SafeToOpen fails at $\a-$, holds over $[\a,\b)$ and fails at $\b$.

\end{enumerate} 
\end{lemma}

\begin{proof}\ 
\begin{enumerate}
\item
Use part 3 of Local SafeToOpen Lemma and the fact that there are only
finitely many tracks. 

\item Use parts 1 and 2 of Local SafeToOpen Lemma.

\item Use parts 1 and 2 of Local SafeToOpen Lemma.

\item Suppose that SafeToOpen holds at $t$.  By parts 1 and 2 of Local
SafeToOpen Lemma, for every track $x$, $t$ belongs to an interval $[\a_x
<\b_x)$ such that $s(x)$ fails at $\a_x-$, holds over $[\a_x,\b_x)$ and
fails at $\b_x$.  The desired $\a =\max_x\a_x$, and the desired $\b
=\min_x\b_x$.\qed
\end{enumerate}
\end{proof}

\begin{lemma}[(Dir Lemma)]
Suppose that $[\a,b)$ is a maximal positive interval of SafeToOpen.
\begin{enumerate}

\item Dir = close at $\a$.

\item Dir = open over $(\a,\b]$ and at $\b+$.

\end{enumerate}
\end{lemma}

\begin{proof}\ 
\begin{enumerate}
\item By Global SafeToOpen Lemma, some $\TS(x)$ becomes empty at $t$.  Fix
such an $x$ and let $0 = t_0 < t_1 < t_2 <\ldots$ be the significant
moments of $\TS(x)$.  Then $\a = t_{3i+3}$ for some $i$.  By Three Rules
Corollary, $\SD(x)$ fires at $t_{3i+1} +\WT$ setting Dir to close.  By
Local SafeToOpen Lemma, $s(x)$ fails over $I = (t_{3i+1} +\WT, 
t_{3i+3}]$.  Hence SafeToOpen fails over $I$ and therefore every $\SC(y)$
is disabled over $I$.  Thus Dir remains close over $I$.

\item
By 1, SignalOpen fires at $\a$ setting Dir to open.  By part 5 of
Local SafeToOpen Lemma, every $\SC(x)$ is disabled over $[\a,\b]$ and at
$\b+$.  Hence Dir remains open over $(\a,\b]$ and at $\b+$.
\qed
\end{enumerate}
\end{proof}

\begin{corollary}[(SignalOpen Corollary)] 
SignalOpen fires exactly when SafeToOpen becomes true. 
SignalOpen fires only when some Track\-Status$(x)$ becomes true. 
\end{corollary}

\begin{proof}
Obvious.
\qed\end{proof}

We have proved some properties of regular runs of our ealgebra $\cA$,
but the question arises if there any regular runs.  Moreover, are
there any regular runs consistent with a given pattern of trains?  The
answer is positive.  In Sect.~8, we will prove that every pattern of
trains gives rise to a regular run and will describe all regular runs
consistent with a given pattern of trains.  

\section{Safety and Liveness}

Recall that we restrict attention to regular runs of our ealgebra $\cA$.

\begin{theorem}[(Safety Theorem)]
The gate is closed whenever a train is in the crossing.  More
formally, $\GS=\closed$ whenever $\TS(x)=\incrossing$ for any $x$.
\end{theorem}

\begin{proof}
Let $t_0 < t_1 <\ldots$ be the significant moments of some track $x$.
Thus, during periods $[t_{3i+2}, t_{3i+3})$, $\TS(x)=\incrossing$.  We
show that $\GS =\closed$ over $[t_{3i+2}, t_{3i+3}]$ and even over
$[t_{3i+1} +\dmin, t_{3i+3}]$.  (Recall that $\dmin\leq t_{3i+2} -
t_{3i+1}\leq\dmax$ and therefore $t_{3i+1} +\dmin\leq t_{3i+2}$.)

By Three Rules Corollary, $\SD(x)$ fires at $t_{3i+1}$ setting
$\DL(x)$ to $\a = t_{3i+1} +\WT$.  If $\Dir_\a = \open$ then $\SC(x)$
fires at $\a$ setting Dir to close; regardless, $\Dir_{\a+} = \close$.
By Local SafeToOpen Lemma, $s(x)$ fails over $I = (\a, t_{3i+3})$.
Hence, over $I$, SafeToOpen fails, SignalOpen is disabled, Dir =
close, and OpenGate is disabled.

By the definition of regular runs, $\GS =\closed$ at some moment $t$
such that $\a < t <\a +\dclose = t_{3i+1} +\WT +\dclose = t_{3i+1}
+\dmin$.  Since OpenGate is disabled over $I$, \GS\ remains closed
over $I$ and therefore over the interval $[t_{3i+1} +\dmin,
t_{3i+3})$.  By Preservation Lemma, $\GS =\closed$ at $t_{3i+3}$.
\qed\end{proof}

Let $\Dclose = \dclose + (\dmax - \dmin) = \dmax - \WT$.  

\begin{theorem}[(Liveness Theorem)] 
Assume $\a +\dopen < \b -\Dclose$.  If the crossing is empty in the
open time interval $(\a,\b)$, then the gate is open in $[\a +\dopen,\b
-\Dclose]$.  More formally, if every $\TS(x)\neq\incrossing$ over
$(\a,\b)$, then $\GS =\opened$ over $[\a +\dopen,\b -\Dclose]$.
\end{theorem}

\begin{proof}
By Deadline Lemma, every $\DL(x)\geq\b -\Dclose >\a +\dopen$ over $(\a,\b)$.
By the definition of SafeToOpen, it holds at $\a$.  If $\Dir_\a =
\close$ then SignalOpen fires at $\a$; in any case $\Dir_{\a+}=\open$.

By Deadline Lemma, every $\DL(x)\geq\b -\Dclose > CT$ over $(\a,\b
-\Dclose)$.  Hence, over $(\a,\b -\Dclose)$, every $\SC(x)$ is
disabled, Dir remains open, and StartClose is disabled.

By the definition of regular runs, $\GS =\opened$ at some moment $t\in(\a,\a
+\dopen)$.  Since StartClose is disabled over $(\a,\b -\Dclose)$, \GS\ remains
opened over $(t,\b -\Dclose)$ and therefore is opened over $[\a +\dopen,\b
-\Dclose)$.  By Preservation Lemma, $\GS =\opened$ at $b -\Dclose$.  
\qed\end{proof}

The next claim shows that, in a sense, Liveness Theorem cannot be
improved. 

\begin{Claim}\ 
\begin{enumerate}

\item Liveness Theorem fails if $\dopen$ is replaced with a smaller
constant.  

\item Liveness Theorem fails if $\Dclose$ is replaced with a smaller
constant.

\end{enumerate}
\end{Claim}

\begin{proof}
The first statement holds because the gate can take time arbitrarily
close to $\dopen$ to open.  The second statement holds for two
reasons.  Recall that $\Dclose = \dclose + (\dmax - \dmin)$.  The term
$(\dmax - \dmin)$ cannot be reduced; to be on the safe side, the
controller must act as if every oncoming train is moving as fast as
possible, even if it is moving as slow as possible.  The term
$\dclose$ cannot be reduced either; the gate can take arbitrarily
short periods of time to close.  Now we give a more detailed proof.

\paragraph{Part 1.}
Given some constant $\copen <\dopen$, we construct a regular run of our
ealgebra $\cA$ and exhibit an open interval $I = (\a,\b)$ such that the
crossing is empty during $I$ but the gate is not opened during a part of
interval $(\a +\copen, \b -\Dclose)$.

We assume that $\dopen,\Dclose < 1$ (just choose the unit of time
appropriately) and that there is only one track.

The traffic.  Only one train goes through the crossing.  It appears at time
$100$, reaches the crossing at time $100 +\dmax$ and leaves the crossing at
time $110 +\dmax$, so that Dir should be changed only twice: set to
close at $100 +\WT$ and set to open at $110 +\dmax$.

The run.  We don't care how quickly the gate closes, but we stipulate that
the time $\Delta$ that the gate takes to open belongs to
$(\copen,\dopen)$.

The interval $I$: $(110 +\dmax, 110 +\dmax +\dopen)$. 

Since the only train leaves the crossing at $110 +\dmax$, the crossing
is empty during $I$.  However the gate takes time $\Delta >\copen$ to
open and thus is not opened during the part $(110 +\dmax +\copen, 110
+\dmax +\Delta)$ of $I$.

\paragraph{Part 2.}
Given some constant $\Cclose <\Dclose$, we construct a regular run of our
ealgebra $\cA$ and exhibit an open interval $I = (\a,\b)$ such that the
crossing is empty during $I$ but the gate is not opened (even closed)
during a part of interval $(\a +\dopen, \b -\Cclose)$.

We assume that $\dopen,\Cclose < 1$, and that there is only one track with
the same traffic pattern as in part 1.

The run.  This time we don't care how quickly the gate opens, but we
stipulate that the time $\Delta$ that the gate takes to close
satisfies the following condition:

\[ 0 < \Delta < \min\{ \dclose, \Dclose -\Cclose \}. \]

The interval $I$ is $(0,100+\dmax)$, so that $\a = 0$ and $\b = 100
+\dmax$.  

Since the only train reaches the crossing at $100+\dmax$, the crossing is
empty during $I$.   The gate is closed by $100 +\WT +\Delta$ and is
closed during the part $(100 +\WT +\Delta, 100 +\WT +(\Dclose -\Cclose))$ of
interval $(\a +\dopen,\b -\Cclose)$.  Let us check that  $(100 +\WT +\Delta,
100 +\WT +(\Dclose -\Cclose)$ is indeed a part of $(\a +\dopen,\b -\Cclose)$.
Clearly, $\a + \dopen < 0 + 1 < 100 + \WT + \Delta$.  Further:
\begin{eqnarray*}
&& 100 +\WT + \Delta \\ 
&<& 100 + \WT + (\Dclose - \Cclose)\\
&=& 100 + (\dmin - \dclose) + [(\dclose + \dmax - \dmin) - \Cclose] =
\b - \Cclose. 
\end{eqnarray*}
\qed\end{proof}

\section{Some Additional Properties}


\begin{theorem}[(Uninterrupted Closing Theorem)]
The closing of the gate is never interrupted.  More formally, if Dir
is set to close at some moment $\a$, then Dir = close over the
interval $I = (\a,\a +\dclose)$.
\end{theorem}

Recall that, by the definition of regular runs, GateStatus = closed
somewhere in $I$ if Dir = close over $I$.

\begin{proof} 
Since Dir is set to close at $\a$, some $\SC(x)$ fires at $\a$.  Fix
such an $x$ and let $t_0 < t_1 <\ldots$ be the significant moments of
track $x$.  By Three Rules Corollary, there is an $i$ such that $\a =
t_{3i+1} +\WT = t_{3i+1} +\dmin -\dclose$.  Then $\a +\dclose =
t_{3i+1} +\dmin\leq t_{3i+2}$. By the definition of regular runs,
$\TS(x) =\coming$ over $I$.  By Deadline Theorem, $\DL(x) =\a$ over
$I$, so that $\CT +\dopen >\CT >\DL(x)$ over $I$.  Because of this
$x$, SafeToOpen fails over $I$ and therefore SignalOpen is disabled
over $I$.  Thus Dir = close over $I$.
\end{proof} 

\begin{theorem}[(Uninterrupted Opening Theorem)]\ 
Suppose $\WT\geq\dopen$; that is, $\dmin\geq\dclose +\dopen$.  Then
the opening of the gate is not interrupted; in other words, if Dir is
set to open at some moment $\a$, then Dir = open over the interval $I
= (\a,\a +\dopen)$.
\end{theorem}

Recall that, by the definition of regular runs, GateStatus = opened
somewhere in $I$ if Dir = open over $I$.  

\begin{proof}
It suffices to prove that every $\SC(x)$ is disabled over $I$.  Pick
any $x$ and let $t_0 < t_1 <\ldots$ be the significant moments of
track $x$.  Since Dir is set to open at $\a$, SignalOpen fires at
$\a$, SafeToOpen holds at $\a$, and $s(x)$ holds at $\a$.  We have two
cases.

\paragraph{Case 1.} $\a +\dopen <\DL(x)_\a <\infty$.  Since $\DL(x)_\a
<\infty$, $\t_{3i+1}<\a\leq t_{3i+3}$ and $\DL(x)_\a = t_{3i+1} +\WT$
for some $i$ (by Deadline Lemma).  We have

\[ \a +\dopen < \DL(x)_\a = t_{3i+1} +\WT < 
   t_{3i+1} +\dmin \leq t_{3i+2} < t_{3i+3}. \]

\noindent
By Deadline Lemma, $\DL(x)$ does not change in $I$, so that CT remains
$<\DL(x)$ in $I$ and therefore $\SC(x)$ is disabled over $I$.

\paragraph{Case 2.} $\a +\dopen\geq\DL_\a(x)$ or $\DL_\a(x) =\infty$.  

We check that $t_{3i}\leq\a\leq t_{3i+1}$ for some $i$.  Indeed, if
$\TS(x)_\a =\empty$ then $t_{3i}\leq\a<t_{3i+1}$ for some $i$.
Suppose that $\TS(x)_\a\neq\empty$.  Since $s(x)$ holds at $a$, $\a
+\dopen <\DL_\a(x)$.  By the condition of Case 2, $\DL(x)_\a =\infty$.
Recall that $\TS(x)\neq\empty$ exactly in intervals $[t_{3i+1},
t_{3i+3}$ and $\DL(x) =\infty$ exactly in periods $(t_{3i},
t_{3i+1}]$.  Thus $\a = t_{3i+1}$ for some $i$.

The first moment after $\a$ that $\SC(x)$ is enabled is $t_{3i+1}
+\WT$.  Thus it suffices to check that $\a +\dopen\leq t_{3i+1} +\WT$.
Since $\dmin \geq\dclose +\dopen$, we have

\[ \a + \dopen \leq t_{3i+1} + \dopen \leq  
t_{3i+1} + (\dmin -\dclose) =  t_{3i+1} + \WT. \qed\]   
\end{proof} 

\begin{corollary}[(Dir and GateStatus Corollary)]
Assume $\dmin\geq\dclose +\dopen$.
\begin{enumerate}

\item If the sequence $\g_1 <\g_2 <\g_3 <\ldots$ of positive significant moments
of Dir is infinite, then the sequence $\d_1 <\d_2 <\d_3 <\ldots$ of positive
significant moments of \GS\ is infinite and each $\d_i\in(\g_i,\g_{i+1})$. 

\item If the positive significant moments of Dir form a finite sequence $\g_1
<\g_2 <\ldots <\g_n$, then the positive significant moments of \GS\ form a
sequence $\d_1 <\d_2 <\ldots <\d_n$ such that $\d_i\in(\g_i,\g_{i+1})$ for all
$i<n$ and $\d_n >\g_n$. 

\end{enumerate}
\end{corollary}

\begin{proof}
We prove only the first claim; the second claim is proved similarly.

Since Dir = open and $\GS =\opened$ initially, \GS\ does not change in
$(0,\g_1)$.  Suppose that we have proved that if $\g_1 <\ldots <\g_j$
are the first $j$ positive significant moments of Dir, then there are
exactly $j-1$ significant moments $\d_1 <\ldots <\d_{j-1}$ of \GS\ in
$(0,g_j]$ and each $\d_i\in(\g_i,\g_{i+1})$.  We restrict attention to
the case when $j$ is even; the case of odd $j$ is similar.  Since $j$
is even, Dir is set to open at $\g_j$.  If $\g_j$ is the last
significant moment of Dir, then the gate will open at some time in
$(\g_j,\g_j +\dopen)$ and will stay open forever after that.
Otherwise, let $k = j + 1$.  By Uninterrupted Opening Theorem, the
gate opens at some moment $\d_j\in(\g_j,\g_k)$.  Since Dir remains
open in $(\d_j,\g_k)$, $\GS =\opened$ holds over $(\d_j,\g_k)$.  By
Preservation Lemma, $\GS =\opened$ at $\g_k$.  \qed\end{proof}

\section{Existence of Regular Runs}

We delayed the existence issue in order to take advantage of Sect.~8.
For simplicity, we restrict attention to an easier but seemingly more
important case when $\dmin \geq\dclose + \dopen$.  The Existence
Theorem and the two Claims proved in this section remain true in the
case $\dmin<\dclose +\dopen$; we provide remarks explaining the
necessary changes.

Let $\U_1 =\U -\{\GS\}$, and $\U_0 =\U_1 -\{\DL,\Dir\}$.  For $i = 0,1$, let
$\U_i^+ =\U_i\cup\{\CT\}$.

\begin{theorem}[(Existence Theorem)] 
Let $P$ be a pre-run of vocabulary $\U_0$ satisfying the train motion
requirement in the definition of regular runs, and let $A$ be an
initial state of $\cA$ consistent with $P(0)$.  There is a regular run
$R$ of $\cA$ which starts with $A$ and agrees with $P$.
\end{theorem}

\begin{proof}  
Let the significant moments of $P$ be $0=\a_0 <\a_1 <\ldots$.  For
simplicity, we consider only the case where this sequence is infinite. 
The case when the sequence is finite is similar.  Our construction
proceeds in two phases.  In the first phase, we construct a run $Q$ of
module\ \Controller\ (that is of the corresponding one-module evolving
algebra of vocabulary $\U_1^+$) consistent with $A$ and $P$.  In the
second phase, we construct the desired $R$ by extending $Q$ to include the
execution of module\ \Gate. \smallskip

\paragraph{Phase 1: Constructing $Q$ from $P$.}  Let $\b_0 <\b_1
<\ldots$ be the sequence that comprises the moments $\a_i$ and the
moments of the form $t +\WT$ where $t$ is a moment when some $\TS(x)$
becomes coming.  By Three Rule and SignalOpen Corollaries, these are
exactly the significant moments of the desired $Q$.  We define the
desired $Q$ by induction on $\b_i$.  It is easy to see that $Q(T)$ is
uniquely defined by its reduct $q(t)$ to $\U_1$.

$Q(0)$ is the appropriate reduct of $A$.  Suppose that $Q$ is defined
over $[0,\b_j]$ and $k=j+1$.  Let $\g$ range over $(\b_j,\b_k)$.  If
\Controller\ does not execute at $\b_j$, define $q(\g) =
q(\b_j)$; otherwise let $q(\g)$ e the state resulting from
executing \Controller\ at $q(\b_j)$.  Define $q(\b_k)$ to agree
with $q(\g)$ at all functions except \TS, where it agrees with
$P(\b_k)$.

Clearly $Q$ is a pre-run.  It is easy to check that $Q$ is a run of
\Controller\ and that \Controller\ is immediate in $Q$. 
\smallskip  

\paragraph{Phase 2: Constructing $R$ from $Q$.}  We construct $R$ by
expanding $Q$ to include \GS.  Let $\g_1 <\g_2 <\ldots$ be the
sequence of significant moments of $Q$ at which Dir changes.  Thus Dir
becomes close at moments $\g_i$ where $i$ is odd, and becomes open at
moments $\g_i$ where $i$ is even.

There are many possible ways of extending $Q$ depending on how long it
takes to perform a given change in \GS.  Chose a sequence
$a_1,a_2,\ldots$ of reals such that (i)~$a_i <\g_{i+1} -\g_i$ and
(ii)~$a_i <\dclose$ if $i$ is odd and $a_i <\dopen$ if $i$ is even.
The idea is that \Gate\ will delay executing OpenGate or CloseGate for time
$a_i$.

The construction proceeds by induction on $\g_i$.  After $i$ steps,
\GS\ will be defined over $[0,g_i]$, and $\GS_{g_i}$ will equal opened
if $i$ is odd and will equal closed otherwise.

Set $\GS =\opened$ over $[0,\g_1]$.  Suppose that \GS\ is defined over
$[0,\g_i]$ and let $j=i+1$.  We consider only the case when $i$ is
even.  The case of odd $i$ is similar.  

By the induction hypothesis, $\GS =\closed$ at $\g_i$.  Since $i$ is even, Dir is
set to open at $\g_i$.  Define $\GS =\closed$ over $(\g_i,\g_i + a_i]$ and
opened over $(\g_i + a_i,\g_j]$.

It is easy to see that $R$ is a regular run of $\cA$.
\qed\end{proof}

Remark.  If the assumption $\dmin\geq\dclose +\dopen$ is removed, Phase 1 of the
construction does not change but Phase 2 becomes more complicated.  After $i$
steps, \GS\ is defined over $[0,g_i]$, and $\GS_{g_i} =\closed$ if $i$ is even;
it cannot be guaranteed that $\GS_{g_i} =\opened$ if $i$ is odd.  The first step
is as above.  For an even $i$, we have three cases.

Case 1: $a_i <\g_j -\g_i$.  Define \GS\ over $(g_i,g_j]$ as in the Existence
Theorem Proof.  

Case 2: $a_i >\g_j -\g_i$.  Define $\GS =\closed$ over $(g_i,g_j]$.

Case 3: $a_i =\g_j -\g_i$.  Define $\GS =\closed$ over $(g_i,g_j]$ as in sub-case 2
but also mark $g_j$ (to indicate that OpenGate should fire at $\g_j$).

For an odd $i$, we have two cases.

Case 1: Either $\GS =\opened$ at $\g_i$ or else $\GS =\closed$ at $g_i$ but $g_i$
is marked.  Define \GS\ over $(g_i,g_j]$ as in the Existence Theorem Proof.

Case 2: $\GS =\closed$ at $\g_i$ and $\g_i$ is not marked.  Ignore $a_i$ and
define $\GS =\closed$ over $(g_i,g_j]$.

\begin{Claim}[(Uniqueness of Control)]
There is only one run of \Controller\ consistent with $A$ and $P$.
\end{Claim}

\begin{proof} 
Intuitively, the claim is true because the construction of $Q$ was
deterministic: we had no choice in determining the significant moments of
$Q$.  More formally, assume by reductio ad absurdum that $Q_1, Q_2$
are runs of \Controller\ consistent with $A$ and $P$ and the set $D = \{t:
Q_1(t)\neq Q_2(t)\}$ is non-empty.  Let $\t=\inf(D)$.  Since both $Q_1$ and
$Q_2$ agree with $A$, $\t>0$.  By the choice of $\t$, $Q_1$ and $Q_2$ agree
over $[0,\t)$.  Since both $Q_1$ and $Q_2$ agree with $A$ and $P$, they can
differ only at internal functions; let $q_1, q_2$ be reductions of $Q_1,
Q_2$ respectively to the internal part of the vocabulary.  By Preservation
Lemma, $q_1$ and $q_2$ coincide at $\t$.  But the values of internal
functions at $\t+$ are completely defined by the state at $t$.  Thus $q_1$
and $q_2$ coincide at $\t+$ and therefore $Q_1, Q_2$ coincide over some
nonempty interval $[\t,\t +\e)$.  This contradicts the definition of $\t$.
\qed\end{proof}

\begin{Claim}[(Universality of Construction)]  
Let $R'$ be any regular run of the ealgebra consistent with $A$ and $P$.  In the
proof of Existence Theorem, the sequence $a_1,a_2,\ldots$ can be chosen in such a
way that the regular run $R$ constructed there coincides with $R'$.
\end{Claim}

\begin{proof}
By Uniqueness of Control Claim, the reducts of $R$ and $R'$ to $\U_1^+$
coincide.  The moments $\g_1<\g_2<\ldots$ when Dir changes in $R$ are exactly the
same moments when Dir changes in $R'$.  We have only to construct appropriate
constants $a_i$.

Let $\d_1 <\d_2 <\dots$ be the significant moments of \GS\ in $R'$. With respect
to Dir and GateStatus Corollary, define $a_i =\d_i -\g_i$.  It is easy to check
that $R = R'$. 
\qed\end{proof}

Remark. If the assumption $\dmin\geq\close +\dopen$ is removed, the proof of
Uniqueness of Control Claim does not change but the proof of Universality of
Construction Claim becomes slightly complicated.  Let $j=i+1$.  For an even $i$,
we have two cases.

Case 1: $\d_i\leq\g_j$.  Define $a_i =\d_i -\g_i$.

Case 2: $\d_i >\g_j$.  In this case $\g_j -\g_i <\dopen$.  The exact value of
$a_i$ is irrelevant; it is only important that $a_i\in(\g_j -\g_i,\dopen)$. 
Choose such an $a_i$ arbitrarily.

For an odd $i$, we also have two cases.

Case 1: In $R'$, either $\GS =\opened$ at $\g_i$ or else $\GS =\closed$ at
$\g_i$ but OpenGate fires at $\g_i$.  Define $a_i =\d_i -\g_i$.   
 
Case 2: In $R'$, $\GS =\closed$ at $\g_i$.   The exact value of $a_i$ is
irrelevant; it is only important that $a_i<\dclose$.  Choose such an $a_i$
arbitrarily.


\begin{thebibliography}{8}

\bibitem{Boerger}
Egon B\"orger, Annotated Bibliography on Evolving Algebras, in
"Specification and Validation Methods", ed. E. B\"orger, Oxford University
Press, 1995, 37--51.

\bibitem{BGR}
Egon B\"orger, Yuri Gurevich and Dean Rosenzweig:
The Bakery Algorithm:  Yet Another Specification and Verification, 
in "Specification and Validation Methods'', ed. E. B\"orger,  
Oxford University Press, 1995.

\bibitem{Guide} 
Yuri Gurevich, ``Evolving Algebra 1993: Lipari
Guide'', in ``Specification and Validation Methods'', Ed.\ E. B\"orger,
Oxford University Press, 1995, 9--36.

\bibitem{tech2} Yuri Gurevich and James K. Huggins, ``The Railroad
Crossing Problem: An Evolving Algebra Solution,'' LITP 95/63, Janvier
1996, Centre National de la Recherche Scientifique Paris, France.

\bibitem{tech1}
Yuri Gurevich, James K. Huggins, and Raghu Mani, ``The Generalized
Railroad Crossing Problem: An Evolving Algebra Based Solution,''
University of Michigan EECS Department Technical Report 
CSE-TR-230-95.

\bibitem{HL} Constance Heitmeyer and Nancy Lynch: The Generalized
Railroad Crossing: A Case Study in Formal Verification of Real-Time
Systems, Proc., Real-Time Systems Symp., San Juan, Puerto Rico, Dec.,
1994, IEEE.

\bibitem{ORS}
Ernst-R\"udiger Olderog, Anders P. Ravn and Jens Ulrik Skakkebaek,
``Refining System Requirements to Program Specifications'', 
to appear.

\bibitem{Milner} 
Robin Milner.  A private discussion, Aug.\ 1994.

\end{thebibliography}
\end{document}